\documentclass[graybox, envcountchap]{svmult}

\usepackage{mathptmx}        
\usepackage{amsmath}
\usepackage{amssymb}
\usepackage{color}
\usepackage{helvet}          
\usepackage{courier}         
\usepackage{dirtree}

\usepackage{makeidx}        
\usepackage{graphicx}        
\usepackage{subfig}
\usepackage{multicol}        
\usepackage[bottom]{footmisc}

\usepackage{hyperref}        
\hypersetup{colorlinks=true,urlcolor=blue}

\usepackage[misc]{ifsym}

\makeindex             

\begin{document}


\title{Resolving the Hubble tension at late times with Dark Energy}
\author{Marco Raveri}

\institute{Marco Raveri (\Letter) \at 
Department of Physics, INFN and INAF at the University of Genova,
Via Dodecaneso 33, 16146 Genova, Italy
\email{marco.raveri@unige.it}}

\maketitle

\abstract{
Within the standard cosmological model, the presence of Dark Energy (DE) is the only structural difference between the early and late times Universe.
While its presence is in full display at late times, it is irrelevant at early times, especially when all other physical ingredients of the standard model are contributing to the formation of CMB anisotropies.
This makes DE a natural candidate to try to solve cosmological tensions between measurements from these two different epochs.
We will analyze how DE presence affects relevant cosmological observables and how it could change them. 
This will allow us to discuss how late-time measurements constrain DE models that aim to resolve the Hubble constant tension, and the achievable performances of these models.
}


\section{Introduction}

In this chapter, we review attempts at solving the Hubble constant tension with non-standard properties of Dark Energy (DE) at late times. 
The empirical evidence for the existence of DE during the late cosmic epochs has been well-established through numerous observational studies, starting from the discovery of cosmic acceleration~\cite{SupernovaSearchTeam:1998fmf, SupernovaCosmologyProject:1998vns}.
Little progress has been made towards knowing its physical nature and properties, that, to this day, remain unknown.
So far, the best description we have for DE is a cosmological constant, $\Lambda$, which is one of the cornerstones of the standard $\Lambda$CDM cosmological model.

DE is a natural candidate to try to explain existing cosmological tensions between the early and late time Universe.
All physical components of the standard $\Lambda$CDM model, but DE, are showcased in the early universe, in particular at recombination. Dark matter, baryons, radiation, and neutrinos are all relevant at times when anisotropies in the Cosmic Microwave Background (CMB) form, and these observations can be used to study their properties. DE, on the other hand, is, at least in the standard model, completely irrelevant at those times.
At late times, in turn, DE becomes relevant and is the only difference between the physical components of the early and late time universe.
This is the empirical motivation to try to tie the physical properties of DE, beyond a cosmological constant, to the resolution of cosmological tensions.

In this chapter, we review these attempts, and the discussion is structured as follows:
we first show, in Sec.~\ref{Sec:DE.Hubble.Const}, how DE enters the cosmological observables that are relevant for the Hubble constant tension;
in Sec.~\ref{Sec:DE.Calibration} we discuss how DE can be used to relieve the tension by changing the calibration of low redshift observables;
in Sec.~\ref{Sec:DE.Expansion} we discuss attempts at solving the Hubble tension by altering the late times expansion history of the Universe with non-standard DE properties.

\section{Dark Energy and the Hubble constant} \label{Sec:DE.Hubble.Const}

The observable universe is characterized by large-scale homogeneity and isotropy, properties well-captured by the flat Friedmann-Lemaître-Robertson-Walker (FLRW) metric with the line element $ds^2 = -dt^2 + a(t)^2d{\bf x}^2$. Here, $a(t)$ represents the scale factor governing the background expansion of the universe, with $t$ being cosmic time. The scale factor is directly related to redshift by $z = 1/a(t) - 1$.

The temporal evolution of the scale factor is governed by the Friedmann equation, given by
\begin{align} \label{Eq:General.Friedman.Equation}
H^2(a) = H_0^2 \left[ \Omega_{\rm r} \, a^{-4} + \Omega_{\rm m} \, a^{-3} + \Omega_{\rm de}(a) \right] \,,
\end{align}
where $H(a) \equiv \dot{a}/{a}$ represents the time-dependent Hubble expansion rate, in cosmic time. In this equation, $H_0$ denotes the present-day Hubble constant, while $\Omega_r$ and $\Omega_m$ correspond to the current relative energy densities of relativistic and non-relativistic particle species, respectively.
The term $\Omega_{\rm de}(a) \equiv \rho_{\rm de}(a) / \rho_{\rm today}^{\rm critical}$ describes the effective density contribution of DE, $\rho_{\rm de}$, relative to present day critical density. 
Note that, from Eq.~\eqref{Eq:General.Friedman.Equation}, at present ($a=1$), the constraint $\Omega_{\rm r} + \Omega_{\rm m} + \Omega_{\rm de} = 1$ has to be satisfied.

In the context of the $\Lambda$CDM model, DE is described with a cosmological constant, $\Lambda$, resulting in a constant value for $\Omega_{\rm de}$. However, in a broader context, $\Omega_{\rm de}(a)$ could encompass the cumulative contribution from all components other than radiation and matter densities. This includes possible modifications to gravity, which would lead to a modified Friedmann equation, as well as the potential existence of a non-zero curvature term, $\Omega_k \, a^{-2}$. 
In all these cases $\Omega_{\rm de}(a)$ becomes explicitly time dependent. A crucial test for the standard $\Lambda$CDM model involves verifying whether $\Omega_{\rm de}$ remains constant throughout cosmic history. 

Friedman equations written as in Eq.~\eqref{Eq:General.Friedman.Equation} are very general and can in principle describe any density contribution. For this reason, this approach is especially useful to probe empirical discrepancies and span different families of models.

Numerous investigations in literature~\cite{Zhao:2017cud, Escamilla:2023oce} focus on the equation of state of DE, which is given by
\begin{align} \label{Eq:DE.equation.of.state}
w_{\rm de}(a) = \frac{1}{3} \frac{d\ln{\Omega_{\rm de}}}{d\ln{a}}-1 \,,
\end{align}
looking for deviations from the standard model value of $w_{\rm de}=-1$, which characterizes a cosmological constant. 
We note here that this approach significantly restricts the family of models that can be studied in the framework, as Eq.~\eqref{Eq:DE.equation.of.state} is well defined only for positive $\Omega_{\rm de}$.
This constraint is not necessarily satisfied in the context of Modified Gravity (MG) theories~\cite{Clifton:2011jh}, in which $\Omega_{\rm de}$ is an effective quantity - a placeholder - for structural modifications of Friedman equations.
This constraint is also not satisfied by curvature effective density contributions and a negative cosmological constant. 
For these reasons, we phrase, whenever possible, the discussion in this chapter in terms of DE effective density.

The Hubble rate determines the redshift dependence of observable cosmological distances, and in particular luminosity distance that is defined as:
\begin{align} \label{Eq:LuminosityDistance}
D_L(z) \equiv \sqrt{\frac{L}{4\pi S}} = (1+z) \int_0^{z} \frac{dz}{H(z)}
\end{align}
where $L$ and $S$ are the observed luminosity and flux of a source and we have assumed spatial flatness, from here onward.

Luminosity distances in the universe at late times can be measured with Supernovae (SN) observations that, once standardized, give us the apparent magnitude of an event, as a function of luminosity distance:
\begin{align} \label{Eq:Apparent.Magnitudes}
m(z) \equiv 5 \log_{10} \left(\frac{D_L(z)}{10\, pc}\right) + M_0
\end{align}
where $M_0$ is a calibration parameter that, crucially, is the same for all (standardized) events of a given type and is redshift independent.
This procedure is discussed in detail, for a state-of-the-art SN survey, in~\cite{Brout:2022vxf}.
Note that in Eq.~\eqref{Eq:Apparent.Magnitudes}, the value of the Hubble constant enters as an additive parameter and, as such, is completely degenerate with the determination of $M_0$.

Distances can also be inferred, at late times, and in full analogy with the CMB, by measuring the angular size of the Baryon Acoustic Oscillation (BAO) feature in the galaxy correlation function.
The reference physical scale of the BAO feature is the sound horizon, $r_s$, of the Thompson scattering coupled photo-baryon fluid, at the time when baryons stopped being dragged by photons. 
Crucially, at late times, this quantity does not evolve with redshift and is fully determined by pre-recombination physics. At a given redshift this scale projects on a feature at an angular separation of:
\begin{align} \label{Eq:BAO.angle}
\theta_{\rm BAO} = \frac{r_s}{D_A(z)}
\end{align}
where $D_A(z)$ is the redshift-dependent angular diameter distance, related to luminosity distance by $D_L = (1+z)^2 D_A$.
The value of the Hubble constant enters multiplicatively in Eq.~\eqref{Eq:BAO.angle}, and, as such, is completely degenerate with the determination of the sound horizon.

\begin{figure}[tbph!]
\includegraphics[width=\columnwidth]{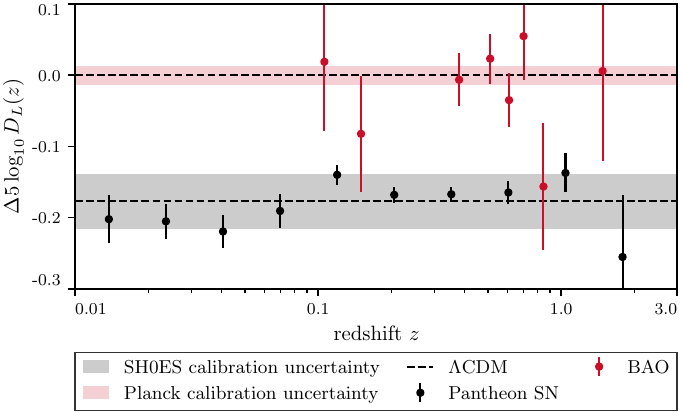}
\caption{\label{fig:shoes} 
The luminosity distance residuals with respect to the $\Lambda$CDM model best fit to CMB observations from Planck~\cite{Planck:2019nip}.
BAO~\cite{eBOSS:2020yzd, Ross:2014qpa, 2011MNRAS.416.3017B} data points, in red, are calibrated by the Planck determination of the physical scale of the sound horizon in the $\Lambda$CDM model.
Pantheon SN~\cite{Pan-STARRS1:2017jku} data points, in black, are calibrated by the SH0ES~\cite{Riess:2020fzl} measurement of absolute SN magnitude, $M_{0}$.
The upper colored band shows the uncertainty in the sound horizon calibration while the lower band shows the errors in the measurements of $M_{0}$.
The separation between the two bands shows in full display the Hubble tension. 
The gap between the data points of the two data sets shows the difficulty of solving the tension with late times modifications to the expansion history:
SN and BAO data overlap in redshift, but, once calibrated, do not overlap in distance space.
Figure adapted from~\cite{Pogosian:2021mcs}. 
}
\end{figure}

Both SN and BAO measurements have to be calibrated, in different ways, to yield distance measurements.
This is what we show in Fig.~\ref{fig:shoes} where we can see: SN measurements from the Pantheon compilation~\cite{Pan-STARRS1:2017jku}, with the redshift binning used in~\cite{Pogosian:2021mcs, Raveri:2021dbu}, calibrated by the measurement of $M_0$ trough the distance ladder, from~\cite{Riess:2020fzl}, which is often phrased as a measurement of $H_0$;
BAO measurements from the eBOSS~\cite{eBOSS:2020yzd}, MGS~\cite{Ross:2014qpa} and 6dF~\cite{2011MNRAS.416.3017B} surveys, calibrated by the measurement of the physical scale of the sound horizon, inferred from CMB observations of the Planck satellite~\cite{Planck:2019nip}.
For both data sets, we show luminosity distance residuals, with respect to the CMB best fit $\Lambda$CDM cosmological model for visual clarity. BAO angular diameter distances are converted to luminosity distances using the effective redshift of the survey tomographic bins.
The gray/red bands show the uncertainty in the calibration of the two data sets respectively and the difference between the two showcases the Hubble tension.
Fig.~\ref{fig:shoes} illustrates the core issue in addressing the Hubble constant tension at late times.
The calibrated SN and BAO data sets share a common redshift range, but their distance measurements do not coincide.
Any given cosmological model, that preserves the behavior of either calibrator, gives a unique distance-redshift relation - is a line in Fig.~\ref{fig:shoes} - and cannot be in two different positions at the same time.

For this reason achieving a complete resolution of the Hubble tension at late times seems unattainable, although several models have been proposed to, at least, alleviate the tension.

A crucial aspect highlighting the challenge of fully addressing the Hubble constant tension at late times arises from the joint use of $H_0/M_0$ measurements, SN, BAO, and CMB data. 
In the context of this chapter, we assume that all four datasets have no residual systematic effects when we discuss DE model performances at solving the tension.
It might be possible to achieve a better resolution of the Hubble constant tension by excluding some parts of these data sets. 

\section{Dark Energy and low redshift calibration} \label{Sec:DE.Calibration}

In this Section, we review attempts at solving the Hubble constant tension by using DE to alter the low redshift calibration of distances.

This calibration is usually reported as a constraint on $H_0$ and, in the standard approach, the local value of $H_0$ is obtained using the SN Ia magnitude-redshift relation, as in Eq.~\ref{Eq:LuminosityDistance}, calibrated with Cepheid variables~\cite{Riess:2021jrx}, with Tip of the Red Giant Branch (TRGB) measurements~\cite{Uddin:2023iob}, and others, see~\cite{Riess:2023egm} for a review.
To strike a balanced tradeoff between sensitivity to expansion history, at high redshift, and contamination from peculiar velocities, at low redshift, only a subset of SN Ia is used in this determination of $H_0$. For example, in~\cite{Riess:2021jrx}, only SN with $z > 0.023$ and $z < 0.15$ are used.
Across this narrow redshift range the sensitivity to the shape of the distance-redshift relation is limited and, in the standard analysis of~\cite{Riess:2021jrx}, it is fit with a template, by fixing the value of the deceleration parameter.

The $H_0$ calibration has then some residual dependency on the low redshift expansion rate and its impact on the determination of the Hubble constant was studied in detail in~\cite{Dhawan:2020xmp}. Several models altering the low redshift expansion history were studied, with an emphasis on DE models.
None of these models were found to give significant changes to the inferred value of $H_0$, with a maximum difference of $\Delta H_0 = 0.47$ Km s$ ^{-1}$ Mpc$ ^{-1}$, or a shift of $0.6\%$, with an expansion history constrained to be very close to the $\Lambda$CDM one.

In retrospect, this conclusion is compatible with residuals in Fig.~\ref{fig:shoes}. The SN calibrator sample includes only the first few low redshift data points and, once the best fitting $\Lambda$CDM expansion history is subtracted, there is no indication of redshift-dependent residuals that could significantly change the estimate of the residuals mean.

We mention here the special case of a very late transition in the DE density, as was studied in~\cite{Benevento:2020fev}. If such a transition occurs very close to the present epoch (i.e. $z < 0.1$ or lower) it would go unnoticed in observations at higher redshifts, like the CMB~\cite{2009PhRvD..80f7301M}. 
For these types of models it is crucial to treat the $H_0$ measurement properly, as a constraint on SN calibration and in~\cite{Raveri:2019gdp, Benevento:2020fev} we discussed how to convert back a constraint on $H_0$ from calibrated SN to a constraint on $M_0$.
A late DE transition, before the first SN at $z<0.01$ can increase the local expansion rate, pushing $H_0$ to values compatible with local measurements of~\cite{Riess:2020fzl} and even higher.
However, just as in void scenarios, the problem with such a solution is that raising $H_0$ does not resolve the origin of the Hubble tension, as shown in Fig~\ref{fig:shoes}, which resides in the calibration of SN in the Hubble flow. In other words, at redshift lower than the Hubble flow SN sample, $z < 0.023$, the expansion history is never relevant as the distance-redshift relation is never used in the fit to derive the value of the Hubble constant, which is effectively extrapolated to $z=0$.

Another, radically different, way of changing the SN Ia calibration is to change the behavior of a step in the distance ladder calibration, especially Cepheids, by altering gravity through DE, as discussed in~\cite{Desmond:2019ygn}.
A modification of gravity through the introduction of a fifth force is a common outcome of theories that couple the degree of freedom of DE with matter, propose novel interactions between objects, or attempt to dynamically explain the nature of dark energy~\cite{Clifton:2011jh, Koyama:2015vza, Joyce:2016vqv, Burrage:2017qrf}.
Given the absence of any observed fifth force in tests of gravity within the solar system, these theories require a mechanism to maintain cosmological relevance while remaining hidden within the solar system.
The concept of screening - which involves weakening the strength of the fifth force in regions of strong gravitational fields - was initially introduced to ensure the consistency of scalar-tensor theories of gravity with tests of the equivalence principle, the inverse square law, and post-Newtonian gravity tests within the confines of the solar system~\cite{Khoury:2010xi, Jain:2010ka}. 
Known screening mechanisms include: chameleon~\cite{Khoury:2003aq, Khoury:2003rn}, symmetron~\cite{Hinterbichler:2010es}, dilaton~\cite{Brax:2010gi}, K-mouflage~\cite{Babichev:2009ee} and Vainshtein~\cite{Vainshtein:1972sx} screening.

Since screening mechanisms generally depend on the local environment they could alter the period-luminosity relation of Cepheids in different host galaxies. In particular, this could happen between the Milky Way and N4258 and Cepheids in the SN Ia calibrator sample, changing the inferred value of the Hubble constant.
In~\cite{Desmond:2019ygn}, the authors assume that the Cepheids used for calibrating the period-luminosity relationship are subject to screening, and compute the expected change in $H_0$, as a function of the residual strength of the fifth force.
Overall changes are not sufficient to fully resolve the Hubble tension: 
models in which the cores of Cepheids are unscreened (leading to increased luminosities) bring the results of the distance ladder closer to, at most, 2-3$\sigma$ from the Planck value when local variations in the effective gravitational constant, $\Delta G / G$ are about 5\%;
models that exclusively unscreen Cepheid envelopes (like chameleon models) require a higher 30-40\% variations in the gravitational constant to achieve a similar level of improvement. 

\section{Dark Energy and the expansion history} \label{Sec:DE.Expansion}

In this Section, we review attempts at solving the Hubble constant tension by using DE to change the expansion history at low to intermediate redshifts. 
This means in particular that we consider changes in the expansion history at $z>0.01$, which allows us to treat the calibration of SN Ia as a constraint on the Hubble constant.

As previously discussed, all these models incur in the no-go argument shown in Sec~\ref{Sec:DE.Hubble.Const} so we generally expect to be able to somewhat relieve the Hubble tension, without resolving it fully.

\subsection{Models for the DE equation of state}

The simplest DE model we can consider, called wCDM, has a constant equation of state, $w_{\rm de} = w_0$, in analogy with matter and radiation components. 
In this case, the value of the equation of state fixes the time dependence of the DE density as $\Omega_{\rm de} \propto a^{-3(1+w_0)}$. 
To increase the expansion rate at low redshifts, with respect to constant $\Omega_{\rm de}$, we need $\Omega_{\rm de}$ to increase in time, rather than diluting, and hence we would need $w_0 < -1$.
As discussed in~\cite{DiValentino:2021izs}, because of the CMB geometric degeneracy, this might look like an empirically viable model when considering CMB-only measurements.
Once the degeneracy is broken though, by the use of either SN or BAO data, the model cannot solve the Hubble constant tension, leaving it qualitatively unchanged.

Similar conclusions apply to other phenomenological models for $w_{\rm de}$, as reviewed in~\cite{DiValentino:2021izs}.
If we consider the combination of Planck 2018, SN and BAO measurements, as in Fig.~\ref{fig:shoes}, the baseline $\Lambda$CDM value for the Hubble constant is, $H_0 = 67.4 \pm 0.8$ Km s$^{-1}$ Mpc$^{-1}$, in $\sim 4 \sigma$ tension with results of $H_0 = 73.04 \pm 1.04$ in~\cite{Riess:2021jrx}.
The wCDM model achieves, when fit to the same data combination $H_0 = 68.34 \pm 0.82$ Km s$^{-1}$ Mpc$^{-1}$, as reported in~\cite{DiValentino:2021izs}, which is a $\sim 0.5\sigma$ reduction in the tension.

We could consider as an extension of the wCDM model the Chevallier - Polarski - Linder parameterization (CPL)~\cite{Chevallier:2000qy, Linder:2002et} for the equation of state, in which $w_{\rm de} = w_0 + w_a (1-a)$. 
When fitting our benchmark combination of data this model yields $H_0 = 68.35 \pm 0.84$ Km s$^{-1}$ Mpc$^{-1}$, which is virtually identical to the wCDM case.
Other parametrizations have been tested in~\cite{DiValentino:2021izs}, in particular: the Jassal- Bagla-Padmanabhan parameterization (JBP)~\cite{Jassal:2005qc} model;
the Logarithmic DE equation of state proposed by~\cite{Efstathiou:1999tm};
the BA parameterization put forward by Barboza and Alcaniz~\cite{Barboza:2008rh}.
All these models would not qualitatively change the results with respect to the wCDM model. 
Some of these models might slightly change quantitatively the tension, by either slightly shifting the central value of the $H_0$ estimate or inflating the error bars of the final estimate. None would be on target and we refer the reader to the discussion in~\cite{DiValentino:2021izs} to gauge different model performances.

In this section, we have discussed only phenomenological approaches and not concrete physical model constructions.
On the one hand, it is always possible to build a model that is physically viable and reproduces any given expansion history~\cite{Frusciante:2019xia}.
On the other hand, considering specific models would impose requirements of physical viability on phenomenological quantities~\cite{Peirone:2017lgi}, hence restricting the available parameter space for solutions and necessarily degrading the performances of a model, towards the solution of the Hubble tension.
We refer the reader to~\cite{Schoneberg:2021qvd, DiValentino:2021izs} for an overview of DE model constraints in the context of the Hubble constant tension.

\subsection{Reconstructed Dark Energy models}

While we have seen several models that alleviate the Hubble constant tension, it is natural to ask what are the ultimate performances that can be achieved by DE models at this task.
This requires going beyond testing specific model realizations or simple parametric forms for $w_{\rm de}(z)$ and estimating $\Omega_{\rm de}(z)$ directly from data.
Indeed, employing simpler constructions can introduce a bias into the results and potentially lead to the loss of important information about ways in which tensions can be relieved.

Multiple strategies exist for the non-parametric reconstruction of cosmological functions like $w_{\rm de}(z)$ or $\Omega_{\rm de}(z)$. 
Widely used techniques encompass discretizing the functions at multiple redshifts, adopting Gaussian Processes (GP)~\cite{Shafieloo:2012ht,Seikel:2012uu,Gerardi:2019obr}, and employing the correlated prior method~\cite{Crittenden:2011aa, Zhao:2012aw, Zhao:2017cud, Wang:2018fng}.
Employing a binning strategy, where the function is assumed to be constant and independent within each bin or smoothly interpolated between redshift nodes, leads to results that rely on an unphysical implicit assumption of smoothness. Opting for a limited number of bins could potentially introduce bias into the reconstruction while increasing their number risks fitting noise features. 
On the other hand, the GP approach doesn't impose restrictions on bin numbers but enforces Gaussianity of the space where it is performed. 
This space, incorporates a Gaussian prior that correlates the function across neighboring redshifts. The selection of the GP prior is fundamentally empirical, devoid of any connection to a physical theory. Moreover, the prior's parameters are typically marginalized, which can obscure the Bayesian interpretation of the resulting reconstruction.
The correlated prior method similarly introduces a correlation between neighboring redshifts, but unlike the GP approach, it employs a fixed prior covariance matrix derived from theoretical considerations. This explicit prior enables a clear articulation of the extent to which the data improves upon the prior and facilitates computation of Bayesian evidence that can be compared to that of the $\Lambda$CDM model, as discussed in~\cite{Raveri:2021dbu}.

In this section, we show what happens to the Hubble constant tension when we opt for approaches 1 and 3, following~\cite{Pogosian:2021mcs, Raveri:2021dbu}.
We do not pursue the GP strategy as we want to reconstruct $\Omega_{\rm de}(z)$. 
Distance measurements, for both SN and BAO, are Gaussian distributed in data space but their measurements are non-linearly related to $\Omega_{\rm de}(z)$, which means that we generally expect the posterior distribution of $\Omega_{\rm de}(z)$ to be non-Gaussian. 
In the reconstruction $\Omega_{\rm de}(z)$ is represented by its value at 11 discrete points (nodes) in redshift, with a cubic spline used to interpolate between them. The first 10 nodes are distributed uniformly in the interval $a \in [1, 0.25]$ while the last node is positioned at $a=0.2$. 
At higher redshift $\Omega_{\rm de}(z)$ was made to approach its $\Lambda$CDM value, even though it is possible to study, with the same framework earlier times deviations from the standard model~\cite{Lin:2018nxe}.

On its own, the cubic spline introduces an implicit smoothing (and correlation) between values of $\Omega_{\rm de}(z)$ at different redshifts, as shown in~\cite{Raveri:2021dbu}.
In addition the time correlation of $\Omega_{\rm de}(z)$ is set from the theoretical correlation matrix obtained in~\cite{Raveri:2017qvt, Espejo:2018hxa} for the general family of Horndeski DE models~\cite{Horndeski:1974wa}.
This theory prior has a much longer correlation time, with respect to the spline model, and acts as a filter, discouraging but not completely prohibiting fast variations in $\Omega_{\rm de}(z)$.
Since the correlation time is much longer than the spline model the theory prior reconstruction is independent of the number of bins used in the reconstruction.

We note that the discussion of the Hubble constant tension, in Sec.~\ref{Sec:DE.Hubble.Const}, relies only on cosmological background arguments. For this reason, we show results for $\Omega_{\rm de}(z)$ and do not discuss the behavior of inhomogeneities beyond the standard model. We refer the reader to~\cite{Pogosian:2021mcs, Raveri:2021dbu} for an in-depth discussion and how it relates to other tensions.

\begin{figure}[tbph!]
\includegraphics[width=\columnwidth]{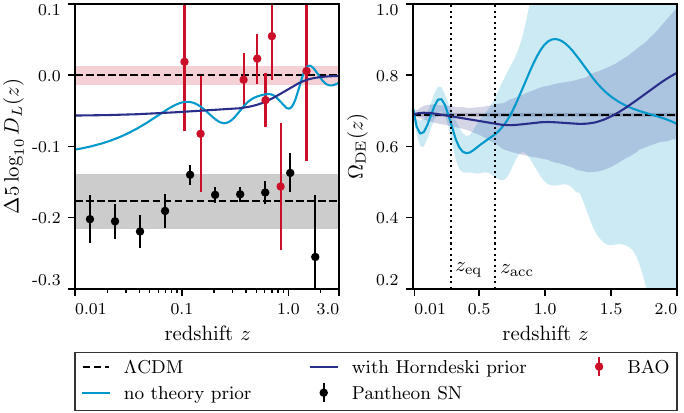}
\caption{\label{fig:reconstruction} 
{\it Left panel:} luminosity distance residuals, as in Fig.~\ref{fig:shoes}, with the two best fitting reconstructed DE models, with and without theoretical (Horndeski) priors.
{\it Right panel:} reconstructed $\Omega_{\rm de}(z)$ for the data sets discussed in this chapter, with and without theory priors.
The bands show the $68\%$ confidence level regions. The two vertical lines show the redshifts of equality between the matter and DE densities, $z_{\rm eq}$, and the beginning of cosmic acceleration, $z_{\rm acc}$, in the best fit $\Lambda$CDM model.
Figure adapted from~\cite{Pogosian:2021mcs}. 
}
\end{figure}

As in~\cite{Pogosian:2021mcs, Raveri:2021dbu} we jointly fit all data discussed in Sec.~\ref{Sec:DE.Hubble.Const}, and the resulting reconstruction of $\Omega_{\rm de}(z)$ is shown in Fig.~\ref{fig:reconstruction}.
The reconstruction is performed with and without theoretical priors and the difference between the two showcases the important role played by the theory priors.
Without theory priors one can see oscillations, in the redshift range $0 < z < 0.6$, in the reconstructed $\Omega_{\rm de}(z)$, which are mainly driven by the scatter of BAO and SN data. These oscillations do not significantly improve the fit and are hence suppressed by the theory prior.
Fig.~\ref{fig:reconstruction} demonstrates the general difficulty of solving the Hubble tension by changing the expansion history at late times. The CMB anchors the expansion rate at high redshift and then the reconstruction struggles to strike a balance between fitting jointly SN and BAO data points.

For the spline reconstruction, which deploys the largest fitting freedom we consider, the inferred value for the Hubble constant, in the reconstructed DE model, is $H_0 = (69.9)\, 69.3 \pm 1.4$ Km s$^{-1}$ Mpc$^{-1}$. In parentheses, hereafter, we have indicated the value for the best-fitting model, which is significantly higher than the mean due to the non-Gaussianity of the posterior that we had anticipated.
If the calibration of SN is included in the fit, to force the reconstruction to struggle to increase the estimate of the Hubble constant, then we obtain $H_0 = (71.34)\, 70.9 \pm 1.4$.
From these two results, we can gauge the extent to which it is possible to relieve the Hubble constant tension at late times. In the first case, the significance is slightly lowered, to about $2 \sigma$, by a slight shift and significantly enlarged error bars. 
In the second case, the significance of the tension is not changed - as the joint dataset contains the SN calibration~\cite{Raveri:2018wln} but it is helpful to gauge the maximum performances that are attainable.
Note that both reconstructions achieve a value of the inferred Hubble constant that is significantly higher than the DE models we discussed in the previous section, because of the much larger fitting freedom.

When we include the DE theoretical priors in the reconstruction the ability of the model to relieve the Hubble constant tension is significantly lowered. When fitting only to CMB, BAO, and SN we can obtain at most $H_0 = (67.7) 67.8^{+1.1}_{-1.4}$ Km s$^{-1}$ Mpc$^{-1}$, a slight improvement over the standard $\Lambda$CDM result of $H_0 = 67.4 \pm 0.8$ Km s$^{-1}$ Mpc$^{-1}$.

\section{Summary}

In this chapter, we have discussed how DE at late times could be invoked to solve the tension over the value of the Hubble constant, between early and late time Universe measurements.
DE is a natural candidate to provide a resolution to this conundrum: its physical properties are unknown; it's the only physical constituent that is irrelevant at early times and important at late times.

We have seen how DE at late times enters into the discussion over the Hubble constant tension, how it influences the cosmological observables that are relevant to the tension, and how this limits the capability of DE models to fully solve it.
Changing the expansion history, or the laws of gravity, at very late times, as we have discussed in Sec.~\ref{Sec:DE.Calibration}, could help alleviate the tension
by changing the very way in which the distance to Hubble flow SN is calibrated.
Changing the expansion history at intermediate redshifts, in the range $0.01 < z < 3$, as we have reviewed in Sec.~\ref{Sec:DE.Expansion}, can also help alleviate the tension. For this family of models, the main limitation to a full tension resolution is that the distance redshift relation is tomographically probed by both SN and BAO observations. Both probes constrain the expansion history, in most of this redshift interval, better than what is needed to solve the tension.
We have seen how this translates to the impossibility of achieving a full tension resolution, even when deploying very general techniques that reconstruct, rather than constrain, DE models at late times.
These approaches are so general that they provide an upper bound to the performances that can be achieved, which in this case are at about $H_0 = (71.34) 70.9 \pm 1.4$, thus solving the tension only halfway.

\begin{acknowledgement}
We thank Levon Pogosian for helpful discussions and comments.
M.R. acknowledges financial support from the INFN InDark initiative.
\end{acknowledgement}
\bibliographystyle{ieeetr}
\bibliography{biblio}

\begin{thebibliography}{10}

\bibitem{SupernovaSearchTeam:1998fmf}
A.~G. Riess {\em et~al.}, ``{Observational evidence from supernovae for an
  accelerating universe and a cosmological constant},'' {\em Astron. J.},
  vol.~116, pp.~1009--1038, 1998.

\bibitem{SupernovaCosmologyProject:1998vns}
S.~Perlmutter {\em et~al.}, ``{Measurements of $\Omega$ and $\Lambda$ from 42
  high redshift supernovae},'' {\em Astrophys. J.}, vol.~517, pp.~565--586,
  1999.

\bibitem{Zhao:2017cud}
G.-B. Zhao {\em et~al.}, ``{Dynamical dark energy in light of the latest
  observations},'' {\em Nature Astron.}, vol.~1, no.~9, pp.~627--632, 2017.

\bibitem{Escamilla:2023oce}
L.~A. Escamilla, W.~Giar\`e, E.~Di~Valentino, R.~C. Nunes, and S.~Vagnozzi,
  ``{The state of the dark energy equation of state circa 2023},'' 7 2023.

\bibitem{Clifton:2011jh}
T.~Clifton, P.~G. Ferreira, A.~Padilla, and C.~Skordis, ``{Modified Gravity and
  Cosmology},'' {\em Phys. Rept.}, vol.~513, pp.~1--189, 2012.

\bibitem{Brout:2022vxf}
D.~Brout {\em et~al.}, ``{The Pantheon+ Analysis: Cosmological Constraints},''
  {\em Astrophys. J.}, vol.~938, no.~2, p.~110, 2022.

\bibitem{Planck:2019nip}
N.~Aghanim {\em et~al.}, ``{Planck 2018 results. V. CMB power spectra and
  likelihoods},'' {\em Astron. Astrophys.}, vol.~641, p.~A5, 2020.

\bibitem{eBOSS:2020yzd}
S.~Alam {\em et~al.}, ``{Completed SDSS-IV extended Baryon Oscillation
  Spectroscopic Survey: Cosmological implications from two decades of
  spectroscopic surveys at the Apache Point Observatory},'' {\em Phys. Rev. D},
  vol.~103, no.~8, p.~083533, 2021.

\bibitem{Ross:2014qpa}
A.~J. Ross, L.~Samushia, C.~Howlett, W.~J. Percival, A.~Burden, and M.~Manera,
  ``{The clustering of the SDSS DR7 main Galaxy sample \textendash{} I. A 4 per
  cent distance measure at $z = 0.15$},'' {\em Mon. Not. Roy. Astron. Soc.},
  vol.~449, no.~1, pp.~835--847, 2015.

\bibitem{2011MNRAS.416.3017B}
F.~{Beutler}, C.~{Blake}, M.~{Colless}, D.~H. {Jones}, L.~{Staveley-Smith},
  L.~{Campbell}, Q.~{Parker}, W.~{Saunders}, and F.~{Watson}, ``{The 6dF Galaxy
  Survey: baryon acoustic oscillations and the local Hubble constant},'' {\em
  Mon. Not. Roy. Astron. Soc.}, vol.~416, pp.~3017--3032, Oct. 2011.

\bibitem{Pan-STARRS1:2017jku}
D.~M. Scolnic {\em et~al.}, ``{The Complete Light-curve Sample of
  Spectroscopically Confirmed SNe Ia from Pan-STARRS1 and Cosmological
  Constraints from the Combined Pantheon Sample},'' {\em Astrophys. J.},
  vol.~859, no.~2, p.~101, 2018.

\bibitem{Riess:2020fzl}
A.~G. Riess, S.~Casertano, W.~Yuan, J.~B. Bowers, L.~Macri, J.~C. Zinn, and
  D.~Scolnic, ``{Cosmic Distances Calibrated to 1\% Precision with Gaia EDR3
  Parallaxes and Hubble Space Telescope Photometry of 75 Milky Way Cepheids
  Confirm Tension with $\Lambda$CDM},'' {\em Astrophys. J. Lett.}, vol.~908,
  no.~1, p.~L6, 2021.

\bibitem{Pogosian:2021mcs}
L.~Pogosian, M.~Raveri, K.~Koyama, M.~Martinelli, A.~Silvestri, G.-B. Zhao,
  J.~Li, S.~Peirone, and A.~Zucca, ``{Imprints of cosmological tensions in
  reconstructed gravity},'' {\em Nature Astron.}, vol.~6, no.~12,
  pp.~1484--1490, 2022.

\bibitem{Raveri:2021dbu}
M.~Raveri, L.~Pogosian, M.~Martinelli, K.~Koyama, A.~Silvestri, G.-B. Zhao,
  J.~Li, S.~Peirone, and A.~Zucca, ``{Principal reconstructed modes of dark
  energy and gravity},'' {\em JCAP}, vol.~02, p.~061, 2023.

\bibitem{Riess:2021jrx}
A.~G. Riess {\em et~al.}, ``{A Comprehensive Measurement of the Local Value of
  the Hubble Constant with 1 km s$^{-1}$ Mpc$^{-1}$ Uncertainty from the Hubble
  Space Telescope and the SH0ES Team},'' {\em Astrophys. J. Lett.}, vol.~934,
  no.~1, p.~L7, 2022.

\bibitem{Uddin:2023iob}
S.~A. Uddin {\em et~al.}, ``{Carnegie Supernova Project-I and -II: Measurements
  of $H_0$ using Cepheid, TRGB, and SBF Distance Calibration to Type Ia
  Supernovae},'' 8 2023.

\bibitem{Riess:2023egm}
A.~G. Riess and L.~Breuval, ``{The Local Value of H$_0$},'' 8 2023.

\bibitem{Dhawan:2020xmp}
S.~Dhawan, D.~Brout, D.~Scolnic, A.~Goobar, A.~G. Riess, and V.~Miranda,
  ``{Cosmological Model Insensitivity of Local $H_0$ from the Cepheid Distance
  Ladder},'' {\em Astrophys. J.}, vol.~894, no.~1, p.~54, 2020.

\bibitem{Benevento:2020fev}
G.~Benevento, W.~Hu, and M.~Raveri, ``{Can Late Dark Energy Transitions Raise
  the Hubble constant?},'' {\em Phys. Rev. D}, vol.~101, no.~10, p.~103517,
  2020.

\bibitem{2009PhRvD..80f7301M}
M.~{Mortonson}, W.~{Hu}, and D.~{Huterer}, ``{Hiding dark energy transitions at
  low redshift},'' {\em Phys. Rev. D}, vol.~80, p.~067301, Sept. 2009.

\bibitem{Raveri:2019gdp}
M.~Raveri, G.~Zacharegkas, and W.~Hu, ``{Quantifying concordance of correlated
  cosmological data sets},'' {\em Phys. Rev. D}, vol.~101, no.~10, p.~103527,
  2020.

\bibitem{Desmond:2019ygn}
H.~Desmond, B.~Jain, and J.~Sakstein, ``{Local resolution of the Hubble
  tension: The impact of screened fifth forces on the cosmic distance
  ladder},'' {\em Phys. Rev. D}, vol.~100, no.~4, p.~043537, 2019.
\newblock [Erratum: Phys.Rev.D 101, 069904 (2020), Erratum: Phys.Rev.D 101,
  129901 (2020)].

\bibitem{Koyama:2015vza}
K.~Koyama, ``{Cosmological Tests of Modified Gravity},'' {\em Rept. Prog.
  Phys.}, vol.~79, no.~4, p.~046902, 2016.

\bibitem{Joyce:2016vqv}
A.~Joyce, L.~Lombriser, and F.~Schmidt, ``{Dark Energy Versus Modified
  Gravity},'' {\em Ann. Rev. Nucl. Part. Sci.}, vol.~66, pp.~95--122, 2016.

\bibitem{Burrage:2017qrf}
C.~Burrage and J.~Sakstein, ``{Tests of Chameleon Gravity},'' {\em Living Rev.
  Rel.}, vol.~21, no.~1, p.~1, 2018.

\bibitem{Khoury:2010xi}
J.~Khoury, ``{Theories of Dark Energy with Screening Mechanisms},'' 11 2010.

\bibitem{Jain:2010ka}
B.~Jain and J.~Khoury, ``{Cosmological Tests of Gravity},'' {\em Annals Phys.},
  vol.~325, pp.~1479--1516, 2010.

\bibitem{Khoury:2003aq}
J.~Khoury and A.~Weltman, ``{Chameleon fields: Awaiting surprises for tests of
  gravity in space},'' {\em Phys. Rev. Lett.}, vol.~93, p.~171104, 2004.

\bibitem{Khoury:2003rn}
J.~Khoury and A.~Weltman, ``{Chameleon cosmology},'' {\em Phys. Rev. D},
  vol.~69, p.~044026, 2004.

\bibitem{Hinterbichler:2010es}
K.~Hinterbichler and J.~Khoury, ``{Symmetron Fields: Screening Long-Range
  Forces Through Local Symmetry Restoration},'' {\em Phys. Rev. Lett.},
  vol.~104, p.~231301, 2010.

\bibitem{Brax:2010gi}
P.~Brax, C.~van~de Bruck, A.-C. Davis, and D.~Shaw, ``{The Dilaton and Modified
  Gravity},'' {\em Phys. Rev. D}, vol.~82, p.~063519, 2010.

\bibitem{Babichev:2009ee}
E.~Babichev, C.~Deffayet, and R.~Ziour, ``{k-Mouflage gravity},'' {\em Int. J.
  Mod. Phys. D}, vol.~18, pp.~2147--2154, 2009.

\bibitem{Vainshtein:1972sx}
A.~I. Vainshtein, ``{To the problem of nonvanishing gravitation mass},'' {\em
  Phys. Lett. B}, vol.~39, pp.~393--394, 1972.

\bibitem{DiValentino:2021izs}
E.~Di~Valentino, O.~Mena, S.~Pan, L.~Visinelli, W.~Yang, A.~Melchiorri, D.~F.
  Mota, A.~G. Riess, and J.~Silk, ``{In the realm of the Hubble
  tension\textemdash{}a review of solutions},'' {\em Class. Quant. Grav.},
  vol.~38, no.~15, p.~153001, 2021.

\bibitem{Chevallier:2000qy}
M.~Chevallier and D.~Polarski, ``{Accelerating universes with scaling dark
  matter},'' {\em Int. J. Mod. Phys. D}, vol.~10, pp.~213--224, 2001.

\bibitem{Linder:2002et}
E.~V. Linder, ``{Exploring the expansion history of the universe},'' {\em Phys.
  Rev. Lett.}, vol.~90, p.~091301, 2003.

\bibitem{Jassal:2005qc}
H.~K. Jassal, J.~S. Bagla, and T.~Padmanabhan, ``{Observational constraints on
  low redshift evolution of dark energy: How consistent are different
  observations?},'' {\em Phys. Rev. D}, vol.~72, p.~103503, 2005.

\bibitem{Efstathiou:1999tm}
G.~Efstathiou, ``{Constraining the equation of state of the universe from
  distant type Ia supernovae and cosmic microwave background anisotropies},''
  {\em Mon. Not. Roy. Astron. Soc.}, vol.~310, pp.~842--850, 1999.

\bibitem{Barboza:2008rh}
E.~M. Barboza, Jr. and J.~S. Alcaniz, ``{A parametric model for dark energy},''
  {\em Phys. Lett. B}, vol.~666, pp.~415--419, 2008.

\bibitem{Frusciante:2019xia}
N.~Frusciante and L.~Perenon, ``{Effective field theory of dark energy: A
  review},'' {\em Phys. Rept.}, vol.~857, pp.~1--63, 2020.

\bibitem{Peirone:2017lgi}
S.~Peirone, M.~Martinelli, M.~Raveri, and A.~Silvestri, ``{Impact of
  theoretical priors in cosmological analyses: the case of single field
  quintessence},'' {\em Phys. Rev. D}, vol.~96, no.~6, p.~063524, 2017.

\bibitem{Schoneberg:2021qvd}
N.~Sch\"oneberg, G.~Franco~Abell\'an, A.~P\'erez~S\'anchez, S.~J. Witte,
  V.~Poulin, and J.~Lesgourgues, ``{The H0 Olympics: A fair ranking of proposed
  models},'' {\em Phys. Rept.}, vol.~984, pp.~1--55, 2022.

\bibitem{Shafieloo:2012ht}
A.~Shafieloo, A.~G. Kim, and E.~V. Linder, ``{Gaussian Process Cosmography},''
  {\em Phys. Rev. D}, vol.~85, p.~123530, 2012.

\bibitem{Seikel:2012uu}
M.~Seikel, C.~Clarkson, and M.~Smith, ``{Reconstruction of dark energy and
  expansion dynamics using Gaussian processes},'' {\em JCAP}, vol.~06, p.~036,
  2012.

\bibitem{Gerardi:2019obr}
F.~Gerardi, M.~Martinelli, and A.~Silvestri, ``{Reconstruction of the Dark
  Energy equation of state from latest data: the impact of theoretical
  priors},'' {\em JCAP}, vol.~07, p.~042, 2019.

\bibitem{Crittenden:2011aa}
R.~G. Crittenden, G.-B. Zhao, L.~Pogosian, L.~Samushia, and X.~Zhang, ``{Fables
  of reconstruction: controlling bias in the dark energy equation of state},''
  {\em JCAP}, vol.~02, p.~048, 2012.

\bibitem{Zhao:2012aw}
G.-B. Zhao, R.~G. Crittenden, L.~Pogosian, and X.~Zhang, ``{Examining the
  evidence for dynamical dark energy},'' {\em Phys. Rev. Lett.}, vol.~109,
  p.~171301, 2012.

\bibitem{Wang:2018fng}
Y.~Wang, L.~Pogosian, G.-B. Zhao, and A.~Zucca, ``{Evolution of dark energy
  reconstructed from the latest observations},'' {\em Astrophys. J. Lett.},
  vol.~869, p.~L8, 2018.

\bibitem{Lin:2018nxe}
M.-X. Lin, M.~Raveri, and W.~Hu, ``{Phenomenology of Modified Gravity at
  Recombination},'' {\em Phys. Rev. D}, vol.~99, no.~4, p.~043514, 2019.

\bibitem{Raveri:2017qvt}
M.~Raveri, P.~Bull, A.~Silvestri, and L.~Pogosian, ``{Priors on the effective
  Dark Energy equation of state in scalar-tensor theories},'' {\em Phys. Rev.
  D}, vol.~96, no.~8, p.~083509, 2017.

\bibitem{Espejo:2018hxa}
J.~Espejo, S.~Peirone, M.~Raveri, K.~Koyama, L.~Pogosian, and A.~Silvestri,
  ``{Phenomenology of Large Scale Structure in scalar-tensor theories: joint
  prior covariance of $w_{\textrm{DE}}$, $\Sigma$ and $\mu$ in Horndeski},''
  {\em Phys. Rev. D}, vol.~99, no.~2, p.~023512, 2019.

\bibitem{Horndeski:1974wa}
G.~W. Horndeski, ``{Second-order scalar-tensor field equations in a
  four-dimensional space},'' {\em Int. J. Theor. Phys.}, vol.~10, pp.~363--384,
  1974.

\bibitem{Raveri:2018wln}
M.~Raveri and W.~Hu, ``{Concordance and Discordance in Cosmology},'' {\em Phys.
  Rev. D}, vol.~99, no.~4, p.~043506, 2019.

\end{thebibliography}


\end{document}